\documentclass[11pt]{article}
\usepackage[dvips]{graphics}
\begin{document}

\newcommand{\be}{\begin{equation}}
\newcommand{\p}{\partial}
\newcommand{\nn}{\nonumber}

\newcommand{\ee}{\end{equation}}
\newcommand{\beqn}{\begin{eqnarray}}
\newcommand{\eeqn}{\end{eqnarray}}
\newcommand{\f}{\frac}
\newcommand{\w}{\wedge}

\begin{titlepage}
\flushright{WIS/08/04-MARCH-DPP}
\flushright{hep-th/0403245}

\vspace{1in}

\begin{center}
\Large
{\bf Probing Maldacena-Nunez in IR with p ${\rm\overline{D3}}$ branes}
\vspace{1in}

\normalsize

\large{ Shesansu Sekhar  Pal }\\

\normalsize
\vspace{.7in}

{\em Weizmann institute of  science,\\
76100 Rehovot, Israel }\\
{\sf shesansu.pal@weizmann.ac.il}
\end{center}

\vspace{1in}

\baselineskip=24pt
\begin{abstract}
We probed Maldacena-Nunez  solution in IR with p coincident anti D3 branes
and found that these probe branes become a fuzzy NS5 brane. Doing the 
dual analysis i.e. from the NS5 brane point of view with the charge of p anti
D3 brane on the world-volume of NS5 brane, we showed that to leading order 
this potential matches with that of p anti D3 branes and the  potential
on the NS5 brane has a stable minima and have also calculated the potential, 
from the NS5 brane point of view, for a small fluctuation along the radial 
direction.


\end{abstract}

\end{titlepage}

\section{Introduction}

In recent years there are several interesting attempts to understand the 
holographic dual of various supergravity background for which the Yang-Mills 
theory will be less supersymmetric. The examples are Klebanov-Strassler
background \cite{ks} and Maldacena-Nunez backgrounds \cite{mn1}. In the 
former case, KS have considered a stack of N D3 branes along with M fractional
D3 branes, which are D5 branes wrapped over a collapsing two cycle, in 
deformed conifold in Type IIB string theory. This supergravity background is 
smooth \footnote{Unlike the Klebanov-Tseytlin background \cite{kt} where there is a naked singularity.}.   
The corresponding  Yang-Mills 
theory is ${\cal N}=1$ supersymmetric with  gauge group $SU(N+M)\times SU(N)$
and possesses many interesting properties like: confinement in IR, 
chiral symmetry 
breaking and duality cascades in UV.  While, in the latter case \cite{mn1},
 MN considers
N coincident NS5 branes, in Type IIB string theory,  wrapped on a two 
sphere, $S^2$ and twisting the normal bundle in a precise way\footnote{The 
spin connection on the two sphere has been set equal to the gauge potential, 
so 
that there will be a covariantly constant Killing spinor \cite{mn, bvs}. } 
such 
that  the corresponding Yang-Mills theory will be ${\cal N}=1$ supersymmetric. 
MN solution is also smooth, shows confinement and breaks $U(1)_R$ symmetry in 
UV. It's been constructed by taking the solution constructed in \cite{cv} 
in seven dimensional gauged supergravity and uplifting the solution to 
10 dimension. The S-dual solution namely the D5 brane wrapped on 
$S^2$ with a three form field strength, $F_3$,  has  also been  
presented in \cite{mn1}. The major difference between KS and MN solution is 
that in the latter  case there are no regular branes but features 
only fractional 
branes whereas in the former, it has  both regular and fractional branes. \\

There is a search for nonsupersymmetric and stable gravitational background 
for which the corresponding gauge theory will also be nonsupersymmetric. 
The study of Kachru et al.\cite{kpv} shows the presence of a nonsupersymmetric  but only classically stable configuration for arbitrarily long lived and 
 quantum mechanically unstable vacua. In this construction, \cite{kpv} have 
considered a stack of $p$ anti D3 branes probing the KS geometry, in 
particular, 
the tip of the deformed conifold. Near to the tip and 
 for $\f{p}{M}\simeq 3\%$ there exists a classically 
stable,  nonsupersymmetric vacua but for $\f{p}{M}\simeq 8\%$ there exists
a marginally stable minimum and for $\f{p}{M} > 8\%$  
corresponds to a stable but supersymmetric vacua, where M is the flux of the
fractional D3 branes. This supersymmetric vacua comes from ``brane/flux 
annihilation'' in which the flux of $H_3$ jumps down by one unit and 
there remains
M-p D3 branes. The mechanism as studied in \cite{kpv} is: when the p anti
D3 branes are being used to probe the KS solution in IR, then these anti branes
become a fuzzy NS5 branes and these NS5 branes from the north-pole perspective
corresponds to M-p D3 branes and there is a fall of $H_3$ flux by one unit.\\

In a recent study of Aharony et al.\cite{osj},  have shown that a 
small deformation 
to Maldacena-Nunez solution corresponds to a stable nonsupersymmetric
background. In the dual theory it corresponds to giving mass to some of 
the scalars. In the gravitational background the stability comes from the 
fact that the original theory had a mass gap and a small deformation can't
change it, as argued in \cite{osj}.\\

In this paper, we want to present the probing of Maldacena-Nunez D5 brane
background by
a stack of $p$ anti D3 branes in IR following the work of \cite{kpv}. The 
analysis  essentially is done using the Born-Infeld and Chern-Simon action
in the S-dual frame in the above stated background. In the S-dual frame, 
we shall see that these anti D3 branes will couple to a six form potential
$B_6$, but  the background has no such potential. Hence, the dynamics of 
these branes are described by the DBI action. Analyzing the DBI action 
in this background implies that these anti branes will be accumulated at 
$r=0$.

First, when $p$ coincident anti D3 branes are put at the IR of MN background it
becomes a fuzzy NS5 brane due to Myers effect and second, studying the 
dynamics of an NS5 brane with the charge of $p$ anti D3 brane on its 
world volume in the MN background reproduces the energy seen by these $p$
anti D3 branes  to leading order in $\f{p}{N}$, where
N is the number of fractional D3 branes of MN background. From the study of the
fuzzy NS5 brane point of view, we found that the potential seen by this brane 
is stable. 

The paper is organized as follows. In section 2, we shall present the 
Maldacena-Nunez solution of N D5 branes wrapped on a two sphere and its IR 
behavior. In section 3, we shall analyze the dynamics of the  p coincident 
anti D3 brane in the IR of MN solution and show the appearance of 
fuzzy 5 brane and in 
section 4, we shall  do the dual analysis i.e. from the NS5 brane point of 
view and show
that the potential energy for the static case matches with the potential 
energy found in section 3. In section 5, we shall add a small fluctuation in 
the radial direction and repeat the analysis of section of 4. 

\section{Maldacena-Nunez background and its r$\rightarrow$ 0 limit}
The supergravity solution  of N number of D5 branes wrapped on  a 2-sphere, $S^2$ is described by the the Maldacena-Nunez background \cite{mn1}. The 
non trivial field content of this 
background are metric, dilaton, $\phi$ and a magnetic three form field 
strength, $F_3$, and the form of the solution in string frame in  
$\alpha^{'}=1$ units are
\footnote{We are following the notation of \cite{mn1,dlm,npr,bertolini}.} 
\be
\label{metric}
ds^2_{10}=e^{\phi}\bigg[ dx^{\mu}dx_{\mu} +N(e^{2h}d\Omega^2_2+ dr^2+\f{1}{4} (\omega^i-A^i)^2)\bigg],
\ee
where  $d\Omega^2_2=d\theta^2+sin^2\theta d\phi^2$. The world volume direction of the branes are along $x_{\mu}, \theta, \phi$. The $\omega^i$'s are the SU(2)
left invariant one forms and satisfy $d\omega^i=-\f{1}{2} \epsilon_{ijk} \omega^j\w \omega^k$ algebra, which essentially parametrises a 3 sphere, $S^3$. The explicit
form of these $\omega^i$'s are
\be
\left(\begin{array}{c} \omega^1\\  
\omega^2\\
\omega^3\end{array}\right) = 
\left( 
\begin{array}{ccc} cos\psi & sin\psi sin\theta_1 & 0 \\
-sin\psi & cos\psi sin\theta_1 & 0\\ 
0 & cos\theta_1 & 1 \end{array}\right) 
\left(\begin{array}{c} d\theta_1\\
d\phi_1\\
d\psi\end{array}\right),
\ee 
with the ranges for these  angles are $0\leq\theta_i\leq\pi, ~~0\leq\phi_i < 2\pi, ~~0\leq\psi < 4\pi$. The one forms $A^i$'s have the following form
\be 
A^1=-a(r) d\theta,\quad A^2=a(r) sin\theta d\phi,\quad A^3=-cos\theta d\phi,
\ee
with 
\beqn
a(r)&=&\f{2r}{sinh2r}\nn \\
e^{2h}&=& r~coth2r-\f{r^2}{sinh^2~2r}-\f{1}{4}\nn \\
e^{-2\phi}&=& e^{-2\phi_0}\f{2e^h}{sinh~2r},
\eeqn
where $\phi_0$ is defined for the value of dilaton at $r=0$. The 3 form 
looks like
\be 
\label{f3}
F_3=-N\Bigg[\f{1}{4}(\omega^1-A^1)\w(\omega^2-A^2)\w(\omega^3-A^3)-\f{1}{4} F^a\w (\omega^a-A^a)\bigg],
\ee
where  $F^a$ is defined as 
\be 
F^a=dA^a+\f{1}{2}\epsilon_{abc}~ A^b\w A^c.
\ee

Explicitly, the components of it are
\be  
F^1=- \f{da}{dr} dr\w d\theta,\quad F^2= \f{da}{dr} sin\theta dr\w d\phi,\quad F^3=(1-a^2) sin\theta d\theta\w d\phi.
\ee
The two form potential associated to this three form field strength, 
$F_3=dC_2$, is
\be 
C_2=\f{N}{4}\Bigg[ \psi(sin\theta d\theta\w d\phi-sin\theta_1 d\theta_1\w d\phi_1)-cos\theta cos\theta_1 d\phi \w d\phi_1-a (d\theta\w \omega^1-sin\theta d\phi\w \omega^2)\bigg].
\ee

The six form potential that follows from the equation of motion of $F_3$, in Einstein frame, i.e. $d(e^{\phi} \star F_3)=0$,  is 
\be 
C_6=dx^0\w dx^1 \w dx^2 \w dx^3 \w {\it C},
\ee
where ${\it C}$ is \cite{npr} 
\beqn 
{\it C}&=&-\f{N}{8}e^{2\phi}\bigg[  [(a^2-1)a^2 e^{-2h}-16e^{2h}] cos\theta d\phi\w dr -(a^2-1) e^{-2h} \omega^3\w dr\nn \\&&+\f{da}{dr}\big( sin\theta d\phi\w \omega^1+d\theta\w \omega^2\big)\Bigg].
\eeqn

Before getting into the discussion of probe dynamics, let us look at first the 
behavior of the solution in $r\rightarrow 0$ limit. In this limit the 
functions $a(r), \phi(r)$ and $ h(r)$ takes the following form
\be
a(r)\rightarrow 1, \quad e^{2h}\rightarrow 0, \quad e^{2\phi}\rightarrow e^{2\phi_0}.
\ee
Hence, the solution in this limit becomes
\beqn
\label{bg} 
ds^2_{10}&\rightarrow & ds^2_7=e^{\phi_0}\bigg [ dx_{\mu}dx^{\mu}+\f{N}{2}d\Omega^2_3\bigg ],\nn \\
F_3 &\rightarrow & -\f{N}{4}(\omega^1-A^1)\w (\omega^2-A^2)\w (\omega^3-A^3)=-\f{N}{2} g^4\w g^3 \w g^5, \nn \\
e^{2\phi}&\rightarrow& e^{2\phi_0}\equiv g^2_s.
\eeqn
where $d\Omega^2_3=\f{1}{2} (g^5)^2+ (g^3)^2+(g^4)^2$ with
\beqn
\label{g_ks}
g^3&=&\f{1}{\sqrt 2}\big [-sin\theta_1 d\phi_1+cos\psi sin\theta_2 d\phi_2-sin\psi d\theta_2 \big ]\nn \\
g^4&=&\f{1}{\sqrt 2}\big [ d\theta_1 +sin\psi sin\theta_2 d\phi_2 +cos\psi d\theta_2\big ]\nn \\
g^5&=& \big [ d\psi+cos\theta_1 d\phi_1+cos\theta_2 d\phi_2 \big].
\eeqn
This is the $S^3$ of \cite{ks}. We get this, with the following identification
of angular coordinates: $\theta_1\rightarrow \theta_2, \phi_1\rightarrow \phi_2, \theta\rightarrow \theta_1, \phi\rightarrow \phi_1, ~\psi\rightarrow \psi$.
The RR three form field strength, $F_3$, has a quantized flux 
around  $S^3$
\be 
\int_{S^3} F_3=4\pi^2 N,
\ee
where the three cycle of $S^3$ is parameterized by $\theta_2=-\theta_1,~~  \phi_2=-\phi_1$ and $\f{\psi}{2}\rightarrow\psi$, which means that  only the 
range of  $\psi$ coordinate is changed 
 and others are same as before \footnote{ With this choice of parametrisation 
the coordinate $\psi$ ranges from [0,2$\pi$], but this range of $\psi$ 
corresponds to the  double cover of $S^3$. Hence, we shall take the 
range of $\psi$ to be [0,$\pi$]. (Thanks to O. Aharony for discussion).
The volume of a standard unit $S^3$ is $2\pi^2$. While this 
parametrisation with the above range of $\psi$ 
corresponds to a sphere of radius r, with $r=\sqrt{2}$. } 
and the $F_3$ in this choice of parametrisation becomes
\be 
F_3=2 N sin\theta_1 sin^2\psi d\theta_1\w d\phi_1\w d\psi.
\ee
\section{Dynamics of p ${\overline{D3}}$ in MN background}
\label{p_d3branes}
As is being shown in \cite{kpv,dkv} that putting anti-D3 branes in KS 
backgrounds  make these anti-D3 branes to feel a radial force, coming from 
gravity and $F_5$,    towards the tip at r=0. Hence, the anti-D3 branes 
will be accumulated at r=0. 

The same  can also be shown to be true for  the MN background, where we are 
probing it with p ${\overline{D3}}$ branes. The dynamics of p coincident 
${\overline{D3}}$, in the S-dual frame,  is described only by the DBI action,
(more about it is described in the next subsection) and is 
\be 
S_{DBI}=-T_3\int STr \Bigg(e^{-\phi} \sqrt{-det(P[G_{\mu\nu}]) ~det(Q^{'i}_j)}\Bigg).
\ee
The non-commutator term in this action is
\be 
-T_3\int STr \Bigg(e^{-\phi} \sqrt{-det(P[G_{\mu\nu}])}\Bigg).
\ee
For the static gauge choice as written below, corresponds to the pull back of 
$G_{\mu\nu}$ as $e^{\phi}\eta_{\mu\nu}$ i.e. 
$P[G_{\mu\nu}]=e^{\phi}\eta_{\mu\nu}$, follows from eq.(\ref{metric}). 
Hence the potential that these ${\overline D3}$ are going to see
 follows from the above action and is
\be 
V\sim  e^{\phi}.
\ee

 \begin{figure}[htb]
\includegraphics{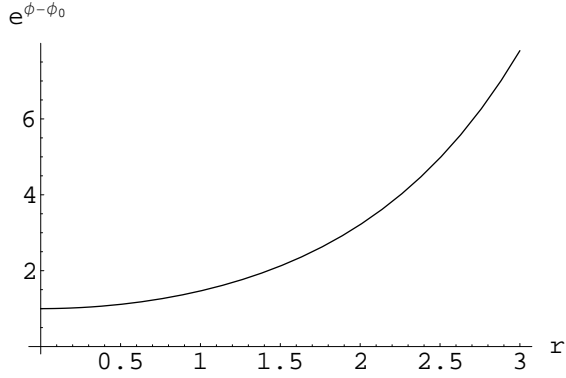}
\caption{The potential, $V\sim  e^{\phi-\phi_0}$  is plotted with  r. 
It follows that the ${\overline D3}$ will be pulled towards $r=0$. }
\label{fig0}
\end{figure} 
The potential is plotted in figure 1 and it is easy to see that the anti 
D3 branes will be accumulated at r=0. 

\subsection{Dielectric effects}
The dynamics of p coincident anti D3 brane is described by Dirac-Born-Infeld
action \cite{myers}
\be
S_{DBI}= -T_3\int STr\Bigg(e^{-\phi} \sqrt{-det(P[E_{ab}+E_{ai}(Q^{-1}-\delta)^{ij}E_{jb}]+\lambda F_{ab})~det(Q^i_j)}\Bigg),
\ee
with \footnote{The dilaton, $\phi$, is not to be confused with one of the angular coordinate that appear in the gauge potential and in the metric of $S^2$.}$E_{ab}=G_{ab}+B_{ab}$ and $Q^i_j=\delta^i_j+i\lambda[\Phi^i,\Phi^k]E_{kj}$. $\lambda=2\pi$, since we are working in $\alpha^{'}=1$ units and the 
Chern-Simon action 
\be 
S_{CS}=-\mu_3\int STr\Bigg(P\bigg[e^{i\lambda i_{\Phi}i_{\Phi}}(\sum C^{(n)}e^B)\bigg]e^{\lambda F}\Bigg).
\ee

The induced metric on the world volume of the anti D3brane, in IR,  
is assumed to be of the following form:
\be 
 ds^2_4=  \eta^{\mu\nu}dx_{\mu}dx_{\nu},
\ee
where we have rescaled the coordinates and assumed the 
static gauge choice i.e. 
$x^{\mu}=\sigma^{\mu}$ and $x^i=\Phi^i$=constants\footnote{From here onwards 
the 
indices $\mu,\nu$ will denote directions parallel to the brane and $i,j$ for 
the transverse directions.}. Hence, there will not be 
any non vanishing $E_{ai}$  term in the DBI action.
The RR two form potential, $C_2$ is non zero only along the direction  
transverse to the world volume of the brane, also we have assumed,  
 that  there will not be any $F_{ab}$ term in the action. Hence, the DBI action
reduces in IR to \footnote{ We have kept $B_2$ term in the action 
but eventually 
it will drop out from eq.(\ref{DBI_action}) for the above stated reason.}
\be 
S_{DBI}=-\f{T_3}{g_s}\int STr \Bigg( \sqrt{-det(P[(\eta+B)_{\mu\nu}]) ~det(Q^i_j)}\Bigg).
\ee
Let us rewrite  the above DBI action in the S-dual frame and is given by
\be 
\label{DBI_action}
-\f{T_3}{g_s}\int STr\Bigg( \sqrt{-det(P[\eta_{\mu\nu}])~det(Q^{'i}_j)}\Bigg),
\ee
with $Q^{'i}_j=\delta^i_j+i\f{\lambda}{g_s}[\Phi^i,\Phi^k](G_{kj}+g_s C_{kj})$.\\

The Chern-Simon action has only one non vanishing term, before taking the
$r\rightarrow 0$ limit i.e.
\be
\label{CS_action} 
S_{CS}=-\mu_3\int STr \Bigg(P\bigg[e^{i\lambda i_{\Phi}i_{\Phi}}C_6\bigg]  \Bigg),
\ee
In this limit the complete action of $p$ coincident anti D3 brane in the 
S-dual frame 
is described by only
 eq. (\ref{DBI_action}) in the background of eq.(\ref{bg}), as in the S-dual 
frame $C_6$ will be replaced by $B_6$ in eq.(\ref{CS_action}). Since, we 
are considering  MN solution described 
by N D5 branes wrapped on $S^2$, which  has no $B_6$ implies that  
in the S-dual frame there will not be Chern-Simon action. The   
fields $\Phi^i$'s
denote the scalar fields in the direction perpendicular to the brane. If we make these fields  to satisfy a non-commutative algebra then the p coincident anti 
D3 branes will represent a 5 dimensional fuzzy brane, following \cite{myers},
 with world-volume topology $R^4\times S^2$. For the case in which the number of anti D3 brane is very small in comparison to the quantized $F_3$ fluxes i.e. $p \ll N$, 
then $\Phi$ remains small relative to the curvature of the spacetime and also 
with respect to the variations in the three form field strength. In this case we can write, following \cite{kpv},  $C_{ij}=\f{4\pi}{3} a F_{ijk}\Phi^k$ with $a=\f{1}{2}$ and the metric in the 
transverse direction may be assumed locally as $G_{ij}=\delta_{ij}$.  The $Q^i_j$ becomes
\be 
Q^i_j=\delta^i_j+\f{2\pi i}{g_s}[\Phi^i,\Phi_j]+i \f{8\pi^2}{3}a F_{kjl}[\Phi^i,\Phi^k]\Phi^l.
\ee
Expanding out the square root term in the Lagrangian of eq. (\ref{DBI_action})  and keeping terms to order $\Phi^4$, we get
\be 
S_{DBI}=-\f{T_3}{g_s}\int  \Bigg(p-i a\f{4 \pi^2}{3} F_{kjl}Tr\bigg([\Phi^k,\Phi^j]\Phi^l\bigg)-\f{\pi^2}{g^2_s} Tr\bigg ([\Phi^i,\Phi^j]^2\bigg)\Bigg),
\ee 
which implies that the potential energy is 
\be
\label{V_p}
V_{p}=-L_{DBI}=\f{T_3}{g_s}  \Bigg(p-ai\f{4 \pi^2}{3} F_{kjl}Tr\bigg([\Phi^k,\Phi^j]\Phi^l\bigg)-\f{\pi^2}{g^2_s} Tr\bigg ([\Phi^i,\Phi^j]^2\bigg)\Bigg).
\ee
The equation of motion that follows from it for $F_{ijk}=2f \epsilon_{ijk}$ is
\be 
\label{eom}
[[\Phi^i,\Phi^j],\Phi^j]-2ig^2_s a f\epsilon_{ijk}[\Phi^j,\Phi^k]=0.
\ee 
The nontrivial solution to eq.(\ref{eom}) is
\be 
[\Phi^i,\Phi^j]=-2ig^2_s a f \epsilon_{ijk}\Phi^k.
\ee
By rescaling the $\Phi^{i}$'s one can realize it as a  $p\times p$ matrix 
representation of SU(2) algebra
\be
\label{su_2} 
[J^i,J^j]=2 i\epsilon_{ijk} J^k,
\ee 
with 
\be
\label{def_phi}
\Phi^i=- a g^2_s J^i.
\ee
The minimum energetic solution would be the p dimensional irreducible 
representation of SU(2) with the second quadratic Casimir as
\be
\label{casimir}
Tr[(J^i)^2]=\f{p}{3}(p^2-1).
\ee
Substituting eq.(\ref{def_phi}) and using  eqs. (\ref{casimir}), (\ref{su_2})
in eq.(\ref{V_p}) gives rise to the potential energy as 
\be
\label{v}
V_p\simeq\f{T_3}{g_s}\Bigg(p- \f{\pi^2}{6} g^6_s 2^4a^4 f^4 p(p^2-1)\Bigg).
\ee
The value to $f$ follows from the normalization of the  three form field 
strength over  $S^3$ and is given by
\be
\label{f}
f=\sqrt\f{1}{N g^3_s}.
\ee
Using eq.(\ref{f}) in eq.(\ref{v}), we get the potential energy as
\be
\label{v_p}
V_p\simeq\f{T_3}{g_s}\Bigg(p-\f{1}{6N^2}\pi^2 p(p^2-1)\Bigg).
\ee
The size of the expanded ${\overline D3}$ branes i.e. fuzzy NS5 brane is 
\be
R=2\pi\sqrt{\sum^3_{i=1}\f{Tr~(\Phi^i)^2}{p}}.
\ee
Using the equation that $\Phi$'s satisfy namely the SU(2) algebra, we get the
radius as
\be
R^2=\f{ \pi^2}{N^2} (p^2-1) R^2_{S^3},
\ee
with $R_{S^3}$ is the radius of $S^3$ and is given by $R^2_{S^3}=Ng_s$.
 
\section{The dual NS5-brane point of view}
The action of an NS5 brane can derived by S-dualising the D5 brane action and 
the dynamics of a D5 brane is governed by the following action \cite{ced}
\beqn
\label{d5_action}
S_{DBI}&=&-T_5\int d^6\sigma e^{-\phi}\sqrt{-det(P[G-B]+2\pi F)},\nn \\
S_{CS}&=& \mu_5\int P[C_6].
\eeqn
In the CS action, we have kept only the non vanishing terms.  
Going over to S-dual frame, the actions become
\beqn
\label{ns5_action}
S_{DBI}&=&-T_5\int d^6 e^{-2\phi}\sigma \sqrt{-det(P[G-e^{\phi} C_2]+2e^{\phi}\pi F)}, \nn \\
S_{CS}&=& \mu_5\int P[B_6]
\eeqn
and this is the action of an NS5 brane. Since, there is no $B_6$ potential in the background implies that the last term will vanish. 
Let us recall that the induced metric in the $r\rightarrow 0$ limit is
\be
\label{ind} 
ds^2_{induced}=dx_{\mu}dx^{\mu}+N g_s [d\psi^2+sin^2\psi d\Omega^2_2],
\ee
where we have rescaled the $x_{\mu}$ coordinates. With this induced metric 
and putting the NS 5brane in the $S^2$ of $S^3$ with radius 
${\sqrt{N g_s}} sin\psi$, the
action of NS 5 brane decomposes, in  IR, to 
\be
\label{det_2}
S=-\f{T_5}{g^2_s}\int d^6\sigma \sqrt{-det(G_{||})~ det(P[G_{\perp}-g_s C_2+2g_s\pi F])}, 
\ee  
with $G_{||}$ denotes the component of the induced metric  along the $x_{\mu}$
and $G_{\perp}$ denotes the component of the induced metric along the $S^2$. 
In IR, $e^{\phi}\rightarrow e^{\phi_0}\equiv g_s$.
 The U(1) field strength corresponding to the p anti D3
brane is 
\beqn
F_{\theta_1\phi_1}&=&\f{p}{2} sin\theta_1 \nn \\
2\pi \int_{S^2} F_2&=&4\pi^2 p,
\eeqn
the $S^2$ is characterized by $ds^2=d\theta^2_1+sin^2\theta_1 d\phi^2_1$ and
the 2 form potential, $C_2$ is  
\beqn
\label{2nd_det_1}
C_{\theta_1\phi_1}&=& N sin\theta_1(\psi-\f{1}{2} sin2\psi)\nn \\
\int_{S^2} C_2 (\psi)&=&4\pi N (\psi-\f{1}{2}sin2\psi).
\eeqn
Hence, the second determinant  in eq.(\ref{det_2}) can be rewritten as
\be
\label{2nd_det_2}
\int_{S^2} \sqrt{det(G_{\perp}+2\pi g_s F - g_s C_2)}= 4 N\pi g_s \Bigg[ sin^4{\psi}+\bigg( \pi \f{p}{N}-\psi+\f{1}{2}sin2\psi\bigg)^2\Bigg]^{\f{1}{2}}
\ee
Using eq.(\ref{ind}) and eq.(\ref{2nd_det_2}) in eq. ({\ref{det_2}}), we 
get the DBI action as 
\be
S=\int d^4\sigma L(\psi),
\ee
with 
\be 
L(\psi)=-\f{T_5}{g_s} 4 N\pi \Bigg[ sin^4{\psi}+\bigg( \pi \f{p}{N}-\psi+\f{1}{2}sin2\psi\bigg)^2\Bigg]^{\f{1}{2}}
\ee
In the case that we are interested in i.e. the static case, the  potential 
energy  that follows from the above Lagrangian is
\be
\label{potential}
V_{eff}(\psi)=-L(\psi).
\ee  
Expanding it for small values of $\psi$, we get
\be
V_{eff}(\psi)\simeq \f{4 \pi^2 T_5 N}{g_s}\bigg(\f{p}{N}-\f{2}{3\pi}\psi^3+\f{N}{2\pi^2 p}\psi^4\bigg),
\ee
this has a minimum at $\psi_{min}=\f{\pi p}{N}$, and the potential energy at 
this minimum becomes
\be
V_{eff}(\psi_{min})= \f{4 \pi^2 T_5 p}{g_s}\bigg( 1-\f{1}{6N^2} \pi^2 p^2\bigg).
\ee
This is the potential seen by p ${\overline D3}$ branes, as calculated in 
eq.(\ref{v_p}). Where we have followed the usual relation between the 
tension of a D5 brane and with the tension of a D3 brane i.e. $4\pi^2 T_5=T_3$.
The potential i.e. eq.(\ref{potential}) is plotted in figure 2.
\begin{figure}[htb]
\includegraphics{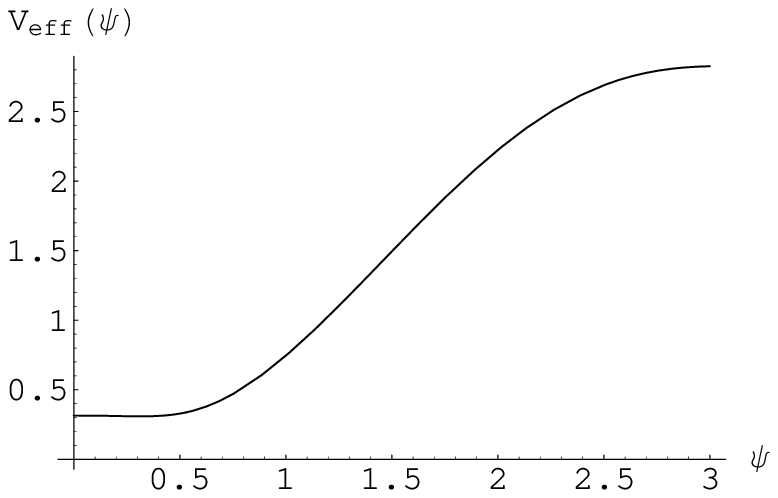}
\includegraphics{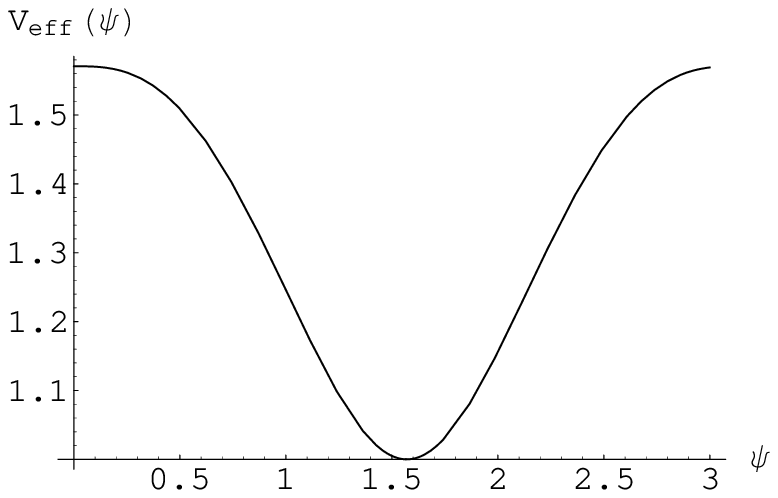}
\caption{The $V_{eff}(\psi)$ i.e. eq.(\ref{potential}) is plotted with $\psi$ for $\f{p}{N}=0.1$ and $\f{p}{N}=0.5$ in units of $\alpha^{'}=1$. It has 
a stable minima.}
\label{figa}
\end{figure}

\section{Fluctuation along the radial direction }
Before we start discussing the behavior of the potential energy of N 
coincident anti D3 branes from the NS5 brane point of view due to fluctuation along the radial direction, let us note the behavior of various fields for 
small $r$. The dilaton
\be
\label{dilaton}
e^{-2\phi}=e^{-2\phi_0} \bigg( 1-\f{8}{9} r^2+\cdots\bigg). 
\ee 
Gauge potential
\beqn
A^1&=&-(1-\f{2}{3}r^2+\cdots)d\theta,\nn \\
A^2&=& (1-\f{2}{3} r^2+\cdots) sin\theta d\phi, \nn \\ 
A^3&=&-cos\theta d\phi.
\eeqn
The geometry in this case with $\theta_1=-\theta_2 ,~~ \phi_1=-\phi_2$,
and $\psi\rightarrow 2\psi$ choice, to quadratic  in $r$, becomes
\beqn 
\label{geometry}
ds^2&\approx&e^{\phi_0}\Bigg(dx^{\mu}dx_{\mu}+N(d\psi^2+sin^2\psi d\Omega^2_2)+\nn \\
& &r^2\big[ \f{4}{9} dx^{\mu}dx_{\mu}+N d\Omega^2_2+ N \f{dr^2}{r^2}+\f{4}{9} N d\psi^2-\f{2}{9} N sin^2\psi d\Omega^2_2\big] \Bigg),
\eeqn
with the expansion of functions $a(r)$ and $h(r)$ as
\beqn
a(r)&=& 1-\f{2}{3} r^2+\cdots,\nn \\
e^{2h}&=&r^2(1-\f{4}{9} r^2+\cdots).
\eeqn
The 3 form RR field strength  i.e. eq.(\ref{f3}) with the above choice of angular coordinates 
$\theta_i$ and $\phi_i$ becomes
\beqn
F_3&=&N \Bigg( (2 sin\theta_1~ sin^2\psi-\f{4}{3}r^2 sin\theta_1~sin^2\psi+\f{2}{3}r^2 sin\theta_1)~d\theta_1\w d\phi_1\w d\psi+\nn \\& &\f{2r}{3} sin2\psi~ sin\theta_1~~ dr\w d\theta_1\w d\phi_1\Bigg)
\eeqn
It is easy to see that the 2 form potential takes the following form
\be
C_2=N\Bigg([(1-\f{2}{3} r^2) sin\theta_1~(\psi-\f{1}{2} sin2\psi)+\f{2}{3}r^2 \psi~ sin\theta_1]~d\theta_1\w d\phi_1\Bigg).\\
\ee
The dynamics of an NS5 brane is described by  
\beqn
\label{NS5_action}
S&=&-T_5\int e^{\phi}\sqrt{-det(P[e^{-\phi}G-C_2]+2\pi F)}\nn \\
&=&-T_5\int e^{-2\phi} \sqrt{-det(P[G-e^{\phi}C_2+2\pi e^{\phi} F])}\nn \\
&\approx& -\f{T_5}{g^2_s}\int (1-\f{8}{9}r^2)\sqrt{-det(P[G-e^{\phi}C_2+2\pi e^{\phi} F])}.
\eeqn  
In the 1st line we have S-dualised the D5 brane action and in the last line the dilaton is expanded  and kept to quadratic in $r$ and as before
 the Chern-Simon action in the S-dual frame vanishes as there is no 
$B_6$ potential to which it can couple, since our background is that of 
D5brane MN solution \cite{mn1}. 

We can rewrite $G-e^{\phi}C_2+2\pi e^{\phi} F$, using eq.(\ref{dilaton}) as
\be
G-g_s(1+\f{4}{9}r^2) C_2+2\pi g_s (1+\f{4}{9} r^2) F,
\ee 
with identifying $g_s=e^{\phi_0}$. Hence, the determinant of 
eq.(\ref{NS5_action}) can be rewritten, using eq.(\ref{geometry}), as
\be 
\label{factor_det}
det(G_{||})det(G_{\perp}-g_s (1+\f{4}{9}r^2) C_2+2\pi g_s (1+\f{4}{9} r^2) F),
\ee
where $G_{||}$ and $G_{\perp}$ denotes components of metric along $x^{\mu}$ 
and along the $\Omega_2$ directions, respectively. Explicitly, $G_{\perp}$ is
given by
\be
G_{\perp}=Ng_s sin^2\psi(1-\f{2}{9}r^2) d\Omega^2_2+ N r^2 g_s d\Omega^2_2.
\ee 
Evaluating the second determinant of eq.(\ref{factor_det}), we get
\beqn
\label{2nd_det}
& &N^2 g^2_s sin^2\theta_1\Bigg(\bigg[(1-\f{2}{9} r^2) sin^2\psi+r^2\bigg]^2+\nn \\& &\bigg[-(1+\f{4}{9} r^2)~(\psi -\f{1}{2}sin~ 2\psi+\f{r^2}{3}sin~ 2\psi)+\f{\pi p}{N}~(1+\f{4}{9} r^2)\bigg]^2\Bigg). 
\eeqn
The number $p$ that appears in the above expression comes 
through the 2 form field 
strength, $F_2$, which corresponds to the number of anti D3 branes that is 
being used to probe the D5 brane background of Maldacena-Nunez, as studied 
in section 
\ref{p_d3branes}. Now integrating it  over the two sphere, $S^2$, and keeping terms to quadratic in r, gives
\beqn
& &\int_{S^2}\sqrt{det(G_{\perp}-g_s(1+\f{4}{9}r^2)C_2+2\pi g_s(1+\f{4}{9} r^2)F)}\nn \\
&=&4\pi Ng_s\Bigg(\sqrt{sin^4\psi+(\pi \f{p}{N}-\psi+\f{1}{2} sin 2\psi)^2}+\nn \\
& &\f{\bigg[2sin^2\psi-\f{4}{9}sin^4\psi+2(\f{4p}{9N}\pi-\f{4}{9}\psi-\f{1}{9}sin 2\psi)(\pi \f{p}{N}-\psi+\f{1}{2}sin 2\psi)\bigg]r^2}{2\sqrt{sin^4\psi+(\pi \f{p}{N}-\psi+\f{1}{2} sin 2\psi)^2}}\Bigg).\nn \\
\eeqn 

Hence the potential that follows from eq.(\ref{NS5_action}), with the induced
metric on the remaining four directions as 
$ds^2_4\approx (1+\f{4}{9} r^2)\eta_{\mu\nu}dx^{\mu}dx^{\nu}$, is
\beqn 
\label{potential_r}
& &V_{eff}(\psi,r)\approx\f{4\pi N T_5}{g_s}\times \Bigg(\sqrt{sin^4\psi+(\pi \f{p}{N}-\psi+\f{1}{2} sin 2\psi)^2}+\nn \\
& &\f{\bigg[sin^2\psi-\f{2}{9}sin^4\psi+(\f{4p}{9N}\pi-\f{4}{9}\psi-\f{1}{9}sin 2\psi)(\pi \f{p}{N}-\psi+\f{1}{2}sin 2\psi)\bigg]r^2}{\sqrt{sin^4\psi+(\pi \f{p}{N}-\psi+\f{1}{2} sin 2\psi)^2}}\Bigg).\nn \\
\eeqn
 \begin{figure}[htb]
\includegraphics{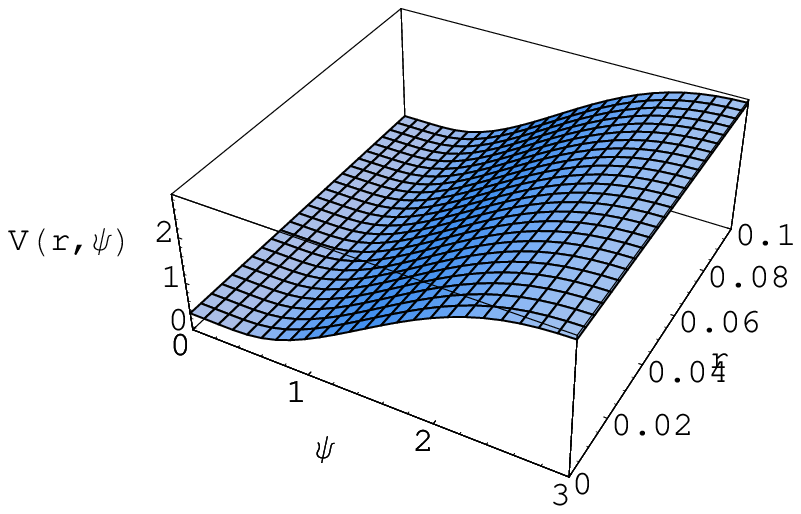}
\includegraphics{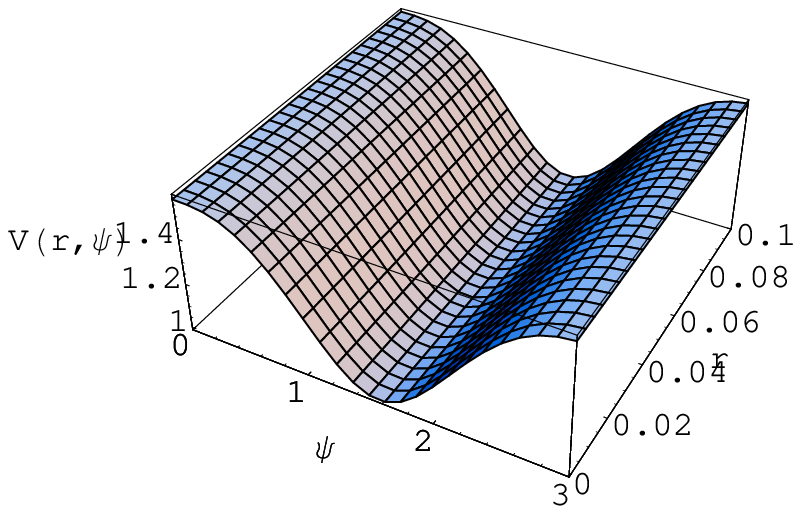}
\includegraphics{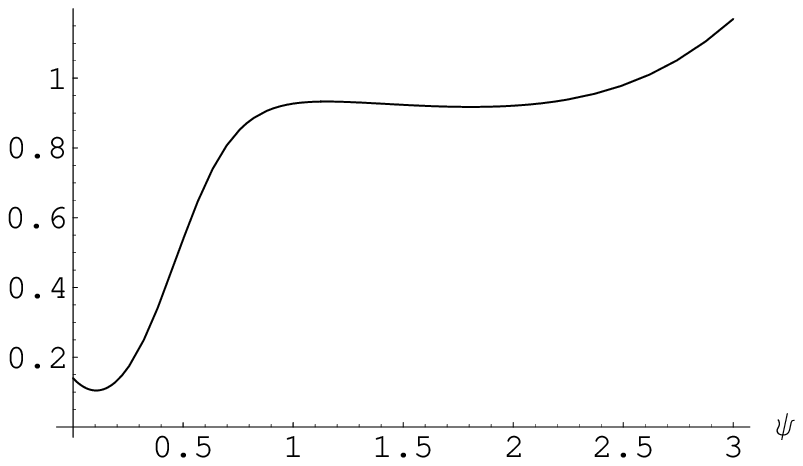}
\caption{The $V_{eff}(\psi, r)$ i.e. eq.(\ref{potential_r}) is plotted with $\psi$ and r for $\f{p}{N}=0.1$ and $\f{p}{N}=0.5$ respectively in units 
of $\alpha^{'}=1$ and the last curve is the coefficient of $r^2$ of 
eq.(\ref{potential_r}) for p/N=0.1.}
\end{figure} 
The potential is plotted in figure 3, for p/N=0.1 and for p/N=0.5. 
It is easy to note that eq.(\ref{potential}) follows in the 
$r\rightarrow 0$ limit. The coefficient of $r^2$ in the 
second term of eq.(\ref{potential_r}) is 
always positive, and is plotted in figure 3 for p/N=0.1, which means that 
the system will always be driven towards the r=0.  
\section{Conclusion}

Probing MN solution of N D5 branes with a stack of p anti D3 branes in IR 
reveals that the potential energy seen by these probe branes is  stable. 
Which means that this probe
brane  have broken all the supersymmetry of MN solution\footnote{Thanks to Ofer Aharony for the discussion.}. We  have also tried to understand some 
kind of geometric 
 transitions between MN backgrounds with  anti D3 branes and the fuzzy NS5 
branes.  However, it would be interesting to understand it by including 
 both the Dp and anti Dp branes and studying it not in the S-dual frame.

The analysis is done as stated above and in the introduction that we have 
considered the probe action of p coincident anti D3 branes and going over to 
the S-dual frame we have the following terms in the DBI action: metric,
two form potential, $C_2$ coming from the three form field strength, 
$F_3=dC_2$,
 and there is no $F_2$, the U(1) field strength. Note that we are keeping 
the term $B_2$ in the DBI action before going to  the S-dual frame even though
we are considering the D5 brane solution of MN, where there is no $B_2$
potential but eventually $C_2$ drops out in the S-dual frame because it has no 
nonzero component along the $x_{\mu}$ directions. The Chern-Simon term drops 
out in the S-dual frame as we do not
have a $B_6$ term in the background. Hence, we left with only the DBI 
action. Taking the $r\rightarrow0$ limit, we analyzed this action in the 
MN background
and found that the p anti D3 brane can become a fuzzy NS5 brane, which 
minimizes the energy. Now, studying the system from the NS5 brane point of 
view,  with the charge corresponding to that of the  p anti D3 branes on the 
world volume
of the NS5 brane, we found that the potential energy matches with that
found in the nonabelian analysis. Note, we construct the NS5 brane action
by starting with the action of a D5 brane which contains metric, $B_2$ and 
$F_2$ 
in the DBI action and a six form potential, $C_6$, in the CS action. Going 
over to 
the S-dual frame, we left with only the DBI action as we do not have any 
$B_6$ term in the background, which can couple to NS5 brane. Hence, in this
case also we left with only the DBI action. As  mentioned above, the leading 
term in the potential matches with that of nonabelian potential of p anti D3 
brane and  the full potential of NS5 brane gives us a minimum which is
stable. Repeating the same NS5 brane analysis but a little bit away from 
the r=0 
and keeping  terms to quadratic in r in potential also makes us to draw 
the same
conclusion as the potential is stable at r=0.     
    

It seems from the analysis we did in section 3 and in 4 that if we replace 
anti D3 branes by D3 branes then the result do not changes since the 
CS action drops out of the probe brane analysis, but we believe that doing 
the analysis not in the S-dual frame might resolve this apparent ambiguities.

Let us note few things if we are not doing the analysis in S-dual frame. 
First the $B_2$ term will drop out from the DBI action of eq.(\ref{d5_action})
as we are  in the 
background of D5 branes of MN but there will be a CS action containing
$i_{\Phi}i_{\Phi} C_6$. Hence, the 
difference will be that of including the CS action and simultaneously 
loosing a term from DBI action. Whereas in the S-dual frame, we have 
no CS action but we have a $C_2$ term in  eq.(\ref{ns5_action}). So, the  
over all square root term in the potential will be changed to a term 
containing square root and the one coming from CS action.  

It would also be interesting  to study the probe brane behavior  in 
IR  both for the MN as well as the KS backgrounds but not going 
to S-dual frame for both Dp and ${\overline Dp}$ branes. Note that the dilaton 
dependence  of a bunch 
of NS5 (D5) branes and NS5 (D5) branes wrapped on a two sphere 
 behave in the same way 
in IR. Hence, it would be interesting also to study these two and to 
see the similarities and differences, if any. 

ACKNOWLEDGMENTS: \\
I would like to thank O. Aharony, M. Berkooz, D. Gepner, S. Kachru, 
S. Kuperstein, Y. Oz,  and 
J. Sonnenschein  for many useful discussions and/or correspondences, 
especially to O. Aharony 
for going through and correcting the manuscript at several places. 
I would also like to thank Feinberg graduate school for providing the 
 fellowship.


\begin{thebibliography}{99}

\bibitem{ks} I.R. Klebanov and M.J. Strassler, ``Supergravity and a Confining Gauge Theory: Duality Cascades and $\chi$SB-resolution of Naked Singularities'', {\em JHEP} {\bf 0008} (2000) 052, hep-th/0007191.

\bibitem{mn1} J.M. Maldacena and C. N${\rm\acute{u}}{\rm\tilde{n}}$ez, ``Towards the large N limit of pure ${\cal N=}1$ super Yang Mills, {\em Phys. Rev. Lett.} {\bf 86}
(2001) 588, hep-th/0008001.

\bibitem{kt} I. R. Klebanov and A. A. Tseytlin, ``Gravity Duals of Supersymmetric $SU(N)\times SU(N+M)$ Gauge Theories,'' hep-th/0002159.

\bibitem{cv} A. H. Chamseddine and M. S. Volkov, ``Non-abelian BPS monopoles in ${\cal N}=4$ gauged supergravity, {\em Phys. Rev. Lett.} {\bf 79} (1997) 3343, hep-th/9707176; ``Non-abelian solutions in ${\cal N}=4$ gauged supergravity and 
leading order string theory'', {\em Phys. Rev.} {\bf D57} (1998) 6242, hep-th/9711181.


\bibitem{mn} J.M. Maldacena and C. N${\rm\acute{u}}{\rm\tilde{n}}$ez, ``Supergravity 
description of field theories on curved manifolds and a no go theorem'', {\em Int. J. Mod. Phys.} {\bf A16} (2001) 822, hep-th/0007018.

\bibitem{bvs} M. Bershadsky, C. Vafa and V. Sadov, ``D-branes and Topological Field Theories", {\em Nucl. Phys.} {\bf B463} (1996) 420, hep-th/9511222.

\bibitem{kpv} S. Kachru, J. Pearson and H. Verlinde, ``Brane/flux annihilation and the string dual of a non-supersymmetric field theory'', {\em JHEP} {\bf 0206} (2002) 021, hep-th/0112197.

\bibitem{osj}O. Aharony, E. Schreiber and J. Sonnenschein, ``Stable non-supersymmetric supergravity solutions from deformations of the Maldacena- N$\rm{\acute{u}}{\rm\tilde{n}}$ez background'', {\em JHEP} {\bf 0204} (2002) 011, hep-th/0201224.
\bibitem{dlm}P. Di Vecchia, A. Lerda and P. Merlatti, ``${\cal N}=1$ and ${\cal N}=2$ Super Yang-Mills theories from wrapped branes'', hep-th/0205204.

\bibitem{bertolini} M. Bertolini, ``Four lectures on the gauge/gravity correspondence,'' hep-th/0303160. 
\bibitem{npr} C. N$\rm{\acute{u}}{\rm\tilde{n}}$ez, $\rm{\acute{A}}$. Paredes and A. Ramallo,  ``Flavoring the gravity dual of ${\cal N}=1$ Yang-Mills with probes'', hep-th/0311201.
\bibitem{myers} R.C. Myers, ``Dielectric-branes'', {\em JHEP} {\bf 12} (1999) 022, hep-th/9910053.
\bibitem{ced} M. Cederwall, A. von Gussich, B.E.W. Nilsson, P. Sundell and A.Westerberg, ``The Dirichlet super-p-branes in Type IIA and IIB supergravity'', {\em Nucl. Phys.} {\bf B490} (1997) 179, hep-th/9611159.

\bibitem{dkv} O. DeWolfe, S. Kachru and H. Verlinde, ``The Giant Inflaton'', 
hep-th/0403123.





\end{thebibliography}
\end{document}